\documentclass{article}
\usepackage[dvipdfmx]{graphicx}
\usepackage{amsmath}
\usepackage{newtxtext,newtxmath}
\usepackage{bm}

\begin{document}

\title{
Fermi-Surface Curvature and Hall Conductivity in Metals
}

\author{
Osamu Narikiyo\thanks{Department of Physics, Kyushu University}}

\date{2020-11-02}

\maketitle

\begin{abstract}
The Tsuji formula which relates the Fermi-surface curvature 
and the weak-field Hall conductivity in metals 
is discussed in Haldane's framework. 
\end{abstract}

\section{Introduction}

The Tsuji formula~\cite{Tsuji} 
relates the Fermi-surface curvature 
and the weak-field Hall conductivity in metals. 
Although it is widely known as a geometrical formula, 
its intelligible derivation has not been reported. 
For example, in the Appendix of an influential textbook~\cite{Hurd} 
the author reported the correspondence with Tsuji 
but it does not seem a derivation. 
I think that not only the derivation but also the meaning of the Tsuji formula 
becomes clear when we employ Haldane's framework.~\cite{Haldane} 
Thus we will discuss the Tsuji formula in Haldane's framework. 

While the Tsuji formula~\cite{Tsuji} 
was derived under the assumption of the cubic symmetry, 
Haldane~\cite{Haldane} tried to eliminate the assumption. 
However, we will criticize the trial. 

This paper is organized as follows. 
In the section 2 we start from the Boltzmann transport theory. 
In the section 3 we focus on the contribution from the Fermi surface. 
We review and criticize Haldane's result in the section 4. 
In the section 5 we analyze the Tsuji formula in Haldane's framework. 
Another geometrical formula by Ong~\cite{Ong} is briefly mentioned in the section 6. 
We give short summary in the last section. 

\section{Boltzmann conductivity}

The weak-field DC Hall conductivity $\sigma^{xy}$ per spin is given by 
\begin{equation}
\sigma^{xy} 
= e^3 \sum_{\bm{k}} 
\left( - {\partial f \over \partial \varepsilon} \right) \ 
l^x  \left( \bm{B} \times \bm{v} \right) \cdot  {\partial \over \partial \bm{k}} \ l^y, 
\label{sigma-xy-1} 
\end{equation}
using the solution of the linearized Boltzmann equation.~\cite{Hurd,Ziman} 
Here the magnetic field $\bm{B}$ is a constant vector. 
The mean free path vector $\bm{l}=(l^x,l^y,l^z)$ is given as $\bm{l}=\tau\bm{v}$ 
where $\bm{v}=(v^x,v^y,v^z)$ is the velocity of the quasi-particle and $\tau$ is its renormalized transport life-time. 
$\partial f / \partial \varepsilon$ is the derivative of the Fermi distribution function $f$ 
with respect to the quasi-particle energy $\varepsilon$. 
Although we have suppressed the argument of $\bm{v}$, $\tau$ and $\varepsilon$, 
they are the functions of the position $\bm{k}=(k^x,k^y,k^z)$ in $\bm{k}$-space. 
The component of the quasi-particle velocity $v^a$ ($a=x,y,z$) is given as 
$v^a = \partial \varepsilon / \partial k^a \equiv \varepsilon_a$. 

It should be noticed that 
Eq. (\ref{sigma-xy-1}) is not the consequence of the relaxation-time approximation 
but the effects of the collision term is fully taken into account 
by renormalizing the life-time.~\cite{KSV} 
Thus Eq. (\ref{sigma-xy-1}) is 
in accordance with the result of the Fermi-liquid theory.~\cite{KY} 

The off-diagonal conductivity, Eq. (\ref{sigma-xy-1}), 
satisfies the Onsager's reciprocal relation: $\sigma_{xy} = - \sigma_{yx}$ with 
\begin{equation}
\sigma^{yx} 
= e^3 \sum_{\bm{k}} 
\left( - {\partial f \over \partial \varepsilon} \right) \ 
l^y  \left( \bm{B} \times \bm{v} \right) \cdot  {\partial \over \partial \bm{k}} \ l^x. 
\label{sigma-yx-1} 
\end{equation}
Actually the integration by parts~\cite{Ziman} for $\sigma_{xy}$ leads to $ - \sigma_{yx}$, 
since  
\begin{equation}
{\partial \over \partial \bm{k}} {\partial f \over \partial \varepsilon} \cdot 
\left( \bm{B} \times \bm{v} \right) = 0,
\label{null-1} 
\end{equation}
and 
\begin{equation}
{\partial \over \partial \bm{k}} \cdot \left( \bm{B} \times \bm{v} \right) = 0. 
\label{null-2} 
\end{equation}
Eq. (\ref{null-1}) and Eq. (\ref{null-2}) 
are the consequences of $\bm{v} = \partial \varepsilon / \partial \bm{k}$. 

In the following 
we switch the factor $(\bm{B}\times\bm{v})\cdot\partial/\partial\bm{k}$ 
to its equivalence $(\bm{v}\times\partial/\partial\bm{k})\cdot\bm{B}$. 
Then Eq. (\ref{sigma-xy-1}) is equivalent to 
\begin{equation}
\sigma^{xy} 
= e^3 \sum_{\bm{k}} 
\left( - {\partial f \over \partial \varepsilon} \right) \ 
l^x  
\left( \bm{v} \times {\partial \over \partial \bm{k}} \right) \cdot \bm{B} \ 
l^y.
\label{sigma-xy-2} 
\end{equation}

If we measure the Hall current in $x$-direction, $J^x$, 
under the electric field in $y$-direction, $\bm{E}=(0,E^y,0)$, 
and the magnetic field in $z$-direction, $\bm{B}=(0,0,B^z)$, 
$ J^x = \sigma^{xzy} B^z E^y $ with 
\begin{equation}
\sigma^{xzy}
= e^3 \sum_{\bm{k}} 
\left( - {\partial f \over \partial \varepsilon} \right) \ 
l^x  
\left( \bm{v} \times {\partial \over \partial \bm{k}} \right)^z 
l^y. 
\label{sigma-xyz} 
\end{equation}
Since $\sigma^{xzy} = - \sigma^{yzx}$, 
we use $e^3 \gamma^{xzy} \equiv ( \sigma_{xzy} - \sigma_{yzx} )/2$ instead of $\sigma^{xzy}$. 
The anti-symmetric property, $\gamma^{xzy} = - \gamma^{yzx}$, is evident in 
\begin{equation}
\gamma^{xzy}
= {1 \over 2} \sum_{\bm{k}} 
\left( - {\partial f \over \partial \varepsilon} \right) \ 
\left[
l^x  
\left( \bm{v} \times {\partial \over \partial \bm{k}} \right)^z 
l^y
-
l^y  
\left( \bm{v} \times {\partial \over \partial \bm{k}} \right)^z 
l^x
\right].
\label{gamma-xyz} 
\end{equation}

Using 
\begin{equation}
\sigma^{xzy}
= e^3 \sum_{\bm{k}} 
\left( - {\partial f \over \partial \varepsilon} \right) \ 
\tau v^x  
\left( v^x {\partial \over \partial k^y} - v^y {\partial \over \partial k^x} \right) 
\tau v^y. 
\label{sigma-xy} 
\end{equation}
we obtain 
\begin{equation}
\gamma^{xzy} 
= {1 \over 2} \sum_{\bm{k}}\left( - {\partial f \over \partial \varepsilon} \right) 
\left( v^x \!,\, v^y \right)
\begin{pmatrix}
M^{-1}_{yy} & -M^{-1}_{yx} \\
-M^{-1}_{xy} & M^{-1}_{xx}
\end{pmatrix}
\begin{pmatrix}
v^x \\
v^y 
\end{pmatrix}
\tau^2 
\label{hat-sigma-xy} 
\end{equation}
where 
\begin{equation}
M_{ab}^{-1} 
\equiv { \partial^2 \varepsilon \over \partial k^a \partial k^b } 
\equiv \varepsilon_{ab},
\label{mass}
\end{equation}
is the effective mass tensor. 
Through the subtraction, $\sigma^{xyz}-\sigma^{yxz}$, 
the derivatives of $\tau$ cancel out. 
This result (\ref{hat-sigma-xy}) is equivalent to 
that of the diagrammatic analysis~\cite{KY} of the linear response of the Fermi liquid. 
Thus the expression for the Hall conductivity 
which contains the derivatives of $\tau$, 
for example (2.53) in Ref. \cite{Hurd}, 
is a bad expression. 

Eq. (\ref{hat-sigma-xy}) is the general result for the Hall conductivity 
so that you have only to estimate it numerically 
if you are not interested in its geometrical interpretation. 

\section{Fermi-surface contribution}

In the case of Fermi degeneracy 
we can estimate $-\partial f / \partial \varepsilon$ by the delta function:
\begin{equation}
\int {\rm d}\bm{k} \left( - {\partial f \over \partial \varepsilon} \right) = \int { {\rm d}S \over |\bm{v}| }, 
\end{equation}
where $ \bm{v} = \partial \varepsilon / \partial \bm{k} $ 
and the integral in the right-hand-side is over the Fermi surface.   
Thus $\gamma^{xzy}$ is determined by the integration over the Fermi surface: 
\begin{equation}
\gamma^{xzy} 
= {1 \over 2} \int { {\rm d}S \over (2\pi)^3 } 
\left( v^x \!,\, v^y \right)
\begin{pmatrix}
M^{-1}_{yy} & -M^{-1}_{yx} \\
-M^{-1}_{xy} & M^{-1}_{xx}
\end{pmatrix}
\begin{pmatrix}
v^x \\
v^y 
\end{pmatrix}
{ \tau^2 \over |\bm{v}| }, 
\label{n-k}
\end{equation}
and reflects the geometry of the Fermi surface. 
Actually it is shown in the following 
that $\gamma^{xzy}$ is a sampling of the local curvatures of the Fermi surface. 

Throughout this paper 
we only consider the contribution from a single sheet of the Fermi surface. 
In the case of multi-sheets 
we should sum the contributions from all the sheets.~\cite{Haldane} 

\section{Haldane's framework}

Haldane used the framework 
\begin{equation}
\sigma^{xy} \equiv e^3 \sum_a \sum_b 
\epsilon^{xya} \gamma_{ab} B^b,
\label{gamma}
\end{equation}
where $\epsilon^{abc}$ is the antisymmetric rank-3 Levi-Civita symbol. 
Since Haldane derived $\gamma_{ab}$ from (\ref{sigma-xy-1}), 
the last equation in p. 2 of Ref. 3, 
$\epsilon^{xyz} \gamma_{zz}$ should be equal to our $\gamma^{xzy}$. 
However, it is different from ours: 
\begin{equation}
\gamma_{zz} 
= {1 \over 2} \int { {\rm d}S \over (2\pi)^3 } 
\left( n^x \!,\, n^y \right)
\begin{pmatrix}
\kappa^{yy} & -\kappa^{yx} \\
-\kappa^{xy} & \kappa^{xx}
\end{pmatrix}
\begin{pmatrix}
n^x \\
n^y 
\end{pmatrix}
l^2,
\label{gamma-Haldane}
\end{equation}
where 
$\bm{n}$ is the unit vector, 
$\bm{n} \equiv \bm{v}/|\bm{v}| \equiv (n^x,n^y,n^z)$, 
which is normal to the Fermi surface, 
$ \kappa^{ab} \equiv \partial n^a / \partial k^b $ 
and $l^2=|\bm{l}|^2=|\bm{v}|^2\tau^2$. 
For example, ours needs only $(\partial v^x / \partial k^y)/|\bm{v}|$ 
but Haldane's needs $\partial(v^x/|\bm{v}|) / \partial k^y$. 
As discussed in the Appendices 
$\kappa^{ab} $ is a key ingredient of the geometrical description 
but its appearance in (\ref{gamma-Haldane}) is incorrect. 
Since it has no geometrical meaning, 
the life-time $\tau$ is safely separated from the other factors 
related to the shape of the Fermi surface as (\ref{n-k}). 
On the other hand, in Eq. (\ref{gamma-Haldane}) 
the mean free path $l$ 
which still contains the geometrical information of the Fermi surface 
is separated. 
Such a separation is a bad strategy and cannot be justified. 
Anyway Eq. (\ref{gamma-Haldane}) is not derived 
from the solution of the Boltzmann equation (\ref{sigma-xy-1}). 

The information on the geometry of the Fermi surface 
can be obtained from the $3 \times 3$ matrix 
\begin{equation}
\begin{pmatrix}
\kappa^{xx} & \kappa^{yx} & \kappa^{zx} \\
\kappa^{xy} & \kappa^{yy} & \kappa^{zy} \\
\kappa^{xz} & \kappa^{yz} & \kappa^{zz} 
\end{pmatrix}.
\label{3x3}
\end{equation}
It can be diagonalized as 
\begin{equation}
\begin{pmatrix}
\kappa_1 & 0 & 0 \\
0 & \kappa_2 & 0 \\
0 & 0 & 0 
\end{pmatrix}.
\end{equation}
The pair of the eigenvalues ($\kappa_1$ and $\kappa_2$) 
is the basis of the geometrical interpretation: 
$ G = \kappa_1 \kappa_2 $ and $ 2H = \kappa_1 + \kappa_2 $ 
where $G$ is the Gaussian curvature and $H$ is the mean curvature. 

Especially the trace $2H$, 
which is independent of the choice of the local coordinate, 
is the target in the next section. 

\section{Tsuji formula in 3D}

Our master equation (\ref{n-k}) is rewritten as 
\begin{equation}
\gamma^{xzy} 
= \int { {\rm d}S \over (2\pi)^3 } 
h_{z} \tau^2, 
\label{Hz}
\end{equation}
where $h_{z}$ is determined by the derivative of the function $\varepsilon$ 
which determines the shape of the Fermi surface: 
\begin{equation}
h_{z} = {1 \over 2 |\bm{v}| }
\left( 
  \varepsilon_x \varepsilon_x \varepsilon_{yy}
+ \varepsilon_y \varepsilon_y \varepsilon_{xx}
- \varepsilon_x \varepsilon_y \varepsilon_{yx}
- \varepsilon_y \varepsilon_x \varepsilon_{xy}
\right),
\label{Gzz}
\end{equation}
with $ \varepsilon_a = \partial \varepsilon / \partial k^a $ and 
$ \varepsilon_{ab} = \partial^2 \varepsilon / \partial k^a \partial k^b $. 
This $h_{z}$ reflects the Fermi-surface geometry but $\tau$ has no geometrical meaning. 

On the other hand, the trace $2H = \kappa^{xx}+\kappa^{yy}+\kappa^{zz}$ 
of the $3 \times 3$ representation (\ref{3x3}) becomes 
\begin{eqnarray}
2H =
{1 \over |\bm{v}|^3 } 
\Big[
\varepsilon_x \varepsilon_x ( \varepsilon_{yy} + \varepsilon_{zz} ) 
&+&
\varepsilon_y \varepsilon_y ( \varepsilon_{zz} + \varepsilon_{xx} )
\nonumber \\
+\ \ \ \varepsilon_z \varepsilon_z ( \varepsilon_{xx} + \varepsilon_{yy} ) 
&-&\varepsilon_x ( \varepsilon_y \varepsilon_{yx} + \varepsilon_z \varepsilon_{zx} ) 
\nonumber \\
-\ \ \ \varepsilon_y ( \varepsilon_x \varepsilon_{xy} + \varepsilon_z \varepsilon_{zy} ) 
&-&\varepsilon_z ( \varepsilon_x \varepsilon_{xz} + \varepsilon_y \varepsilon_{yz} ) 
\Big],
\label{2H}
\end{eqnarray} 
as shown in the Appendix. 
The mean curvature $H$ is 
given by this expression (\ref{2H}) for any shape of the Fermi surface. 

By comparing (\ref{Gzz}) and (\ref{2H}) we see that $h_{z}$ is a piece of $H$. 
By summing three pieces we can construct $H$: $H=(h_{z}+h_{x}+h_{y})/|\bm{v}|^2$ where 
\begin{equation}
h_{x} = {1 \over2 |\bm{v}| }
\left( 
  \varepsilon_y \varepsilon_y \varepsilon_{zz}
+ \varepsilon_z \varepsilon_z \varepsilon_{yy}
- \varepsilon_y \varepsilon_z \varepsilon_{zy}
- \varepsilon_z \varepsilon_y \varepsilon_{yz}
\right),
\label{Gxx}
\end{equation}
and
\begin{equation}
h_{y} = {1 \over2 |\bm{v}| }
\left( 
  \varepsilon_z \varepsilon_z \varepsilon_{xx}
+ \varepsilon_x \varepsilon_x \varepsilon_{zz}
- \varepsilon_z \varepsilon_x \varepsilon_{xz}
- \varepsilon_x \varepsilon_z \varepsilon_{zx}
\right).
\label{Gyy}
\end{equation}

In experiments 
$h_{z}$ is related to the measurement: $ J^x = e^3 \gamma^{xzy} B^z E^y $. 
In the same manner $h_{x}$ and $h_{y}$ are related to the measurements: 
$ J^y = e^3 \gamma^{yxz} B^x E^z $ and $ J^z = e^3 \gamma^{zyx} B^y E^x $ 
where 
\begin{equation}
\gamma^{yxz} 
= \int { {\rm d}S \over (2\pi)^3 } 
h_{x} \tau^2, 
\label{Hx}
\end{equation}
and 
\begin{equation}
\gamma^{zyx} 
= \int { {\rm d}S \over (2\pi)^3 } 
h_{y}\tau^2. 
\label{Hy}
\end{equation}

By summing three experimental results with different configurations we obtain 
\begin{equation}
\gamma^{xzy} + \gamma^{yxz} + \gamma^{zyx} 
= \int { {\rm d}S \over (2\pi)^3 } H l^2, 
\label{any-FS}
\end{equation}
where $l^2=|\bm{v}|^2\tau^2$. 
This relation holds for any shape of the Fermi surface. 

In the case of cubic symmetry 
(\ref{any-FS}) reduces to the Tsuji formula
\begin{eqnarray}
\gamma^{xzy} = \gamma^{yxz} = \gamma^{zyx}
= \int { {\rm d}S \over (2\pi)^3 } { H \over 3 } l^2. 
\label{Tsuji}
\end{eqnarray}

Now we are at the position 
from which we can guess why Haldane expected (\ref{gamma-Haldane}). 
In the case of cubic symmetry 
the integrand in (\ref{Tsuji}) is the product of the scalar $H$ and 
the square of the mean free path $l^2$. 
Haldane introduced a generalized expression by replacing the scalar 
with a tensor keeping $l^2$ in the integrand. 
However, in the case of general symmetry, $l^2$ cannot be factored out. 
We can only factor out $\tau^2$. 

\section{Ong formula in 2D}

Although Haldane~\cite{Haldane} discusses the relation between 2D and 3D formulae, 
we shall discuss the 2D case as a separate issue. 

For simplicity we put $\bm{B}=(0,0,B)$ and set the 2D system in $xy$-plane. 
The 2D version of (\ref{sigma-xy-1}) on the Fermi line is given as  
\begin{equation}
\sigma^{xy} 
= e^3 \int { {\rm d}k_t \over (2\pi)^2 }  
l^x  
\left[  \left( \bm{B}\times\bm{n} \right)\cdot{ \partial \over \partial \bm{k} }  \right]^{z} 
l^y, 
\label{sigma-2D}
\end{equation}
where $ {\rm d} k_t $ is the length along the Fermi line. 
Since $ \bm{B} \times \bm{n} = B \bm{t} $ and $ \bm{t}\cdot(\partial/\partial\bm{k}) = \partial/\partial k_t $, 
$ (\bm{B}\times\bm{n})\cdot(\partial/\partial\bm{k}) = B \partial/\partial k_t $ 
where  $\bm{t}$ is the unit tangent vector along the Fermi line, 
Eq. (\ref{sigma-2D}) is written as 
\begin{equation}
\sigma^{xy} 
= { e^3 B \over (2\pi)^2 } \int l^x  {\rm d} l^y, 
\end{equation}
where $  {\rm d} l^y = ( \partial l^y / \partial k_t ) {\rm d} k_t $. 
After the anti-symmetrization we obtain the Ong formula 
\begin{equation}
{\hat \sigma}^{xy} 
= { e^3 B \over (2\pi)^2 } \int {1 \over 2} \Big[ l^x  {\rm d} l^y - l^y  {\rm d} l^x \Big] 
= { e^3 B \over (2\pi)^2 } \int {1 \over 2} \Big[ \bm{l} \times {\rm d} \bm{l} \Big]^{z}, 
\label{Ong}
\end{equation}
where ${\hat \sigma}^{xy} \equiv ( \sigma_{xy} - \sigma_{yx} ) / 2 $. 
Moreover, Ong~\cite{Ong} discussed the $\lq\lq$Stokes" area in $\bm{l}$-space. 

\section{Summary}

In this paper we have discussed the geometrical formulae 
for the Hall conductivity. 

In 2D the Ong formula is expressed 
by the area in $\bm{l}$-space. 
In 3D the Tsuji formula is expressed 
by the curvature of the Fermi surface in $\bm{k}$-space. 

\appendix

\section{Curvature in differential forms}

In this appendix we review the minimum fundamentals
of a smooth surface $\Sigma$ in 3D Euclidean space. 
(See, for example, \S 4.5 of Ref. \cite{Flanders}.)

Let us choose a point $\bm{x}$ on the surface $\Sigma$ 
and consider the vector $\bm{n}$ normal to $\Sigma$ at $\bm{x}$. 
Then we move to another point $\bm{x}'$ on $\Sigma$ 
and consider the normal vector $\bm{n}'$ there. 
Both $\bm{n}$ and $\bm{n}'$ are unit vectors.  
We assume that the movement is infinitesimally small 
so that both ${\rm d}\bm{x}\equiv\bm{x}'-\bm{x}$ and ${\rm d}\bm{n}\equiv\bm{n}'-\bm{n}$ 
are in the tangent plane at the point $\bm{x}$. 
The normalization $\bm{n}\cdot\bm{n}=1$ leads to ${\rm d}\bm{n}\cdot\bm{n}=0$. 

The vectors in the tangent plane are expanded as 
$ {\rm d}\bm{x} = \sigma_1 \bm{e}_1 + \sigma_2 \bm{e}_2 $ and 
$ {\rm d}\bm{n} = \omega_1 \bm{e}_1 + \omega_2 \bm{e}_2 $ 
where $\bm{e}_1$ and $\bm{e}_2$ are the basis vectors of the tangent plane. 
Here $\sigma_1$, $\sigma_2$, $\omega_1$ and $\omega_2$ are 1-forms. 
The 2-form $\sigma_1\sigma_2$ represents the element of area of $\Sigma$. 
The 2-form $\omega_1\omega_2$ represents the element of area of the unit sphere. 
The Gaussian curvature $K$ is introduced as the magnification factor 
between two areas: $\omega_1\omega_2 = K \sigma_1\sigma_2$. 

Two sets of 1-forms are related by a symmetric matrix ($c=b$) as 
\begin{equation}
\begin{pmatrix}
\omega_1 \\
\omega_2 
\end{pmatrix}
=
\begin{pmatrix}
a & b \\
c & d
\end{pmatrix}
\begin{pmatrix}
\sigma_1 \\
\sigma_2 
\end{pmatrix}.
\end{equation}
The determinant of this matrix is the Gaussian curvature: $K=ad-bc$. 
The trace is related to the mean curvature $H$: $2H=a+d$. 
If this $2 \times 2$ matrix is diagonalized as 
\begin{equation}
\begin{pmatrix}
\kappa_1 & 0 \\
0 & \kappa_2
\end{pmatrix},
\end{equation}
$K = \kappa_1\kappa_2$ and $2H = \kappa_1 + \kappa_2$. 
Here the eigenvalues, $\kappa_1$ and $\kappa_2$, are principal curvatures. 

\section{Components of $2 \times 2$ representation}

In this appendix we calculate the components of the $2 \times 2$ matrix
in the Appendix A explicitly. 
(See, for example, \S 8.2 of Ref. \cite{Flanders}.) 

We give the point on the surface $\Sigma$ as $\bm{x}=(x^1, x^2, u)$ with $u=u(x^1,x^2)$. 
Accordingly ${\rm d}\bm{x} = ({\rm d}x^1, {\rm d}x^2, {\rm d}u) $ 
where ${\rm d}u = p_1{\rm d}x^1 + p_2{\rm d}x^2$ with $ p_i \equiv \partial u/\partial x^i$ ($i=1,2$). 
Introducing the vectors $\bm{t}_1 = (1,0,p_1)$ and $\bm{t}_2 = (0,1,p_2)$ 
the small tangent vector is written as ${\rm d}\bm{x} = \bm{t}_1 {\rm d}x^1 + \bm{t}_2 {\rm d}x^2$. 
The unit normal vector is given 
by $\bm{n} = \bm{w}/|\bm{w}|$ with $\bm{w}=(-p_1,-p_2,1)$. 
It is apparent that $\bm{w}\cdot\bm{t}_1=0$ and $\bm{w}\cdot\bm{t}_2=0$. 

Let us consider the map ${\hat A}$: ${\rm d}\bm{x} \rightarrow {\rm d}\bm{n}$ 
and introduce the $2 \times 2$ representation by 
\begin{equation}
({\rm d}\bm{n}\cdot\bm{t}_i) = \sum_{j=1}^2 a_{ij}  ({\rm d}\bm{x}\cdot\bm{t}_j), 
\end{equation}
where the inner product is defined as 
\begin{equation}
\bm{x}\cdot\bm{t} = \left( x, y, z \right) \cdot \left( t^x, t^y, t^z \right)^{\rm t} 
= \left( x, y, z \right) \cdot 
\begin{pmatrix}
t^x \\ t^y \\ t^z 
\end{pmatrix}
= x t^x + y t^y + z t^z. 
\end{equation}
Since $ {\rm d}\bm{n}\cdot\bm{t}_i = - (1/|\bm{w}|) \sum_j r_{ij} {\rm d}x^j $ and 
$ {\rm d}\bm{x}\cdot\bm{t}_j = \sum_k ( \delta_{jk} + p_j p_k ) {\rm d}x^k $ 
with $ r_{ij} \!=\! \partial^2 u/\partial x^i \partial x^j $, 
$ a_{ij }$ satisfies 
\begin{equation}
\sum_{j=1}^2 a_{ij} \left( \delta_{jk} + p_j p_k \right) = - { 1 \over |\bm{w}| } r_{ik}.
\label{a_ij}
\end{equation}
In terms of Monge's notation ($p \!\!=\!\! \partial u/\partial x,\,\, q \!\!=\!\! \partial u/\partial y,\,\, 
r \!\!=\!\! \partial^2 u/\partial x^2,\,\, s \!\!=\!\! \partial^2 u/\partial x \partial y,\,\, t \!=\! \partial^2 u/\partial y^2$ 
with $x^1 \!\!=\! x$ and $x^2 \!\!=\! y$) 
Eq. (\ref{a_ij}) is written as 
\begin{equation}
{\hat A}
\begin{pmatrix}
1+p^2 & pq \\
pq & 1+q^2
\end{pmatrix}
= - { 1 \over  |\bm{w}| } 
\begin{pmatrix}
r & s \\
s & t
\end{pmatrix}.
\end{equation}
Thus 
\begin{equation}
{\hat A} =  { 1 \over  |\bm{w}|^3 } 
\begin{pmatrix}
pqs - (1+q^2)r & pqr - (1+p^2)s \\
pqt - (1+q^2)s & pqs - (1+p^2)t
\end{pmatrix}.
\end{equation}
The trace is readily obtained as 
\begin{equation}
2H = {\rm trace}\left({\hat A}\right) =  { 1 \over  |\bm{w}|^3 } \Big[ 2pqs - (1+p^2)t - (1+q^2)r \Big].
\end{equation}
After some calculations the determinant is obtained as 
\begin{equation}
G = {\rm det}\left({\hat A}\right) =  { 1 \over  |\bm{w}|^4 } \Big[ rt - s^2 \Big].
\end{equation}

The components $a_{ij}$ are also obtained 
by the derivative of the unit normal vector $ \bm{n} = (n^x, n^y, n^z)$ 
where $\bm{n} = \bm{w}/w$ with $\bm{w}=(-p,-q,1)$ and $w^2 \equiv p^2+q^2+1$.  
Here we put $ p_x \equiv r $, $ q_y \equiv t $ and $ s \equiv p_y = q_x$ 
for the convenience of the calculation. 
If we set the view point at $(0,0,\infty)$, the identification,
$a_{11} = \partial n^x / \partial x$, 
$a_{21} = \partial n^x / \partial y$, 
$a_{12} = \partial n^y / \partial x$ and 
$a_{22} = \partial n^y / \partial y$, 
is naturally understood. 
The results are 
\begin{equation}
a_{11} = { \partial n^x \over \partial x } 
= - {1 \over w} p_x + {p \over w^3} ( pp_x + qq_x ) 
= {1 \over w^3} \Big[ pqs - (1+q^2)r \Big],
\end{equation}
\begin{equation}
a_{21} = { \partial n^x \over \partial y } 
= - {1 \over w} p_y + {p \over w^3} ( pp_y + qq_y ) 
= {1 \over w^3} \Big[ pqt - (1+q^2)s \Big],
\end{equation}
\begin{equation}
a_{12} = { \partial n^y \over \partial x } 
= - {1 \over w} q_x + {q \over w^3} ( pp_x + qq_x ) 
= {1 \over w^3} \Big[ pqr - (1+p^2)s \Big],
\end{equation}
\begin{equation}
a_{22} = { \partial n^y \over \partial y } 
= - {1 \over w} q_y + {q \over w^3} ( pp_y + qq_y ) 
= {1 \over w^3} \Big[ pqs - (1+p^2)t \Big].
\end{equation}
Here we should take care that $ {\rm d}\bm{n} = {\rm d}\bm{x} {\hat A} $.  

\section{\bf{$3 \times 3$} representation}

Here we move from $\bm{x}$-space to $\bm{k}$-space. 
In the Appendix B we have assumed that the $z$-component is given by the function $u(x,y)$ explicitly. 
In the following we assume that the point $\bm{k}=(k^x,k^y,k^z)$ on the Fermi surface 
is given by $\varepsilon(\bm{k})=0$ implicitly. 

The $2 \times 2$ matrix introduced in the Appendix B is written as 
\begin{equation}
\begin{pmatrix}
\kappa^{xx} & \kappa^{yx} \\
\kappa^{xy} & \kappa^{yy}  
\end{pmatrix},
\end{equation}
which is a part of the $3 \times 3$ matrix 
\begin{equation}
\begin{pmatrix}
\kappa^{xx} & \kappa^{yx} & \kappa^{zx} \\
\kappa^{xy} & \kappa^{yy} & \kappa^{zy} \\
\kappa^{xz} & \kappa^{yz} & \kappa^{zz} 
\end{pmatrix},
\end{equation}
where 
\begin{equation}
\kappa^{ab} \equiv { \partial n^a \over \partial k^b },
\label{kappa}
\end{equation}
with $a,b=x,y,z$. 

If the $2 \times 2$ matrix is diagonalized as 
\begin{equation}
\begin{pmatrix}
\kappa_1 & 0 \\
0 & \kappa_2 \\
\end{pmatrix},
\end{equation}
then the $3 \times 3$ matrix is diagonalized as 
\begin{equation}
\begin{pmatrix}
\kappa_1 & 0 & 0 \\
0 & \kappa_2 & 0 \\
0 & 0 & 0 
\end{pmatrix},
\end{equation}
since both $ {\rm d}\bm{n} $ and $ {\rm d}\bm{x} $ are in the tangent plane 
so that the normal vector becomes the eigenvector of the $3 \times 3$ matrix with zero eigenvalue. 
Consequently the trace of the $3 \times 3$ matrix is equal to the trace of $2 \times 2$ matrix. 

The components $\kappa^{ab}$ are 
expressed in terms of the derivative of the quasi-particle energy $\varepsilon$. 
For example, 
\begin{equation}
\kappa^{zz} = { \partial n^z \over \partial k^z } 
=  { \partial \over \partial k^z } { \varepsilon_z \over (\varepsilon_x^2 + \varepsilon_y^2 + \varepsilon_z^2)^{1/2} }
=  {\varepsilon_{zz} \over |\bm{v}|}  - {\varepsilon_z \over |\bm{v}|^3}
   ( \varepsilon_x \varepsilon_{xz} + \varepsilon_y \varepsilon_{yz} + \varepsilon_z \varepsilon_{zz} ),
\label{kappa_zz}
\end{equation}
where $ |\bm{v}|^2 = \varepsilon_x^2 + \varepsilon_y^2 + \varepsilon_z^2 $.
In Eq. (\ref{kappa_zz}) the term $\varepsilon_z \varepsilon_z \varepsilon_{zz}$, 
which is not expected for the off-diagonal conductivity, disappers by the subtraction so that we obtain 
\begin{equation}
\kappa^{zz} = { 1 \over  |\bm{v}|^3 }
\Big[
  ( \varepsilon_x \varepsilon_x + \varepsilon_y \varepsilon_y ) \varepsilon_{zz} 
- \varepsilon_z ( \varepsilon_x \varepsilon_{xz} + \varepsilon_y \varepsilon_{yz} )
\Big].
\label{kappa-diagonal}
\end{equation}
In the same manner we also obtain 
\begin{equation}
\kappa^{xx} = { 1 \over  |\bm{v}|^3 }
\Big[
  ( \varepsilon_y \varepsilon_y + \varepsilon_z \varepsilon_z ) \varepsilon_{xx} 
- \varepsilon_x ( \varepsilon_y \varepsilon_{yx} + \varepsilon_z \varepsilon_{zx} )
\Big],
\label{kappa-diagonal-x}
\end{equation}
and
\begin{equation}
\kappa^{yy} = { 1 \over  |\bm{v}|^3 }
\Big[
  ( \varepsilon_x \varepsilon_x + \varepsilon_z \varepsilon_z ) \varepsilon_{yy} 
- \varepsilon_y ( \varepsilon_x \varepsilon_{xy} + \varepsilon_z \varepsilon_{zy} )
\Big].
\label{kappa-diagonal-y}
\end{equation}

\section{\bf Tsuji and Ong formulae}

The Tsuji formula focuses on the shape of the Fermi surface. 
Thus the integrand of the Tsuji formula becomes $H l^2$ 
where the curvature $H$ represents the geometrical information of the Fermi surface. 
Another factor $l^2$ is determined by the interaction among electrons and is not geometrical. 

On the other hand, the Ong formula focuses on the magnetic flux. 
In the definition of the flux the life-time $\tau$ is involved ($l = \tau v$). 
Symbolically, $\omega_{\rm c} \tau$ represents the flux. 
Such a discussion in 3D is given in the appendix of axXiv:2211.05761v2.

\end{document}